\documentstyle[times,pramana,epsf,floats]{ias}

\pramvolinfo{\hskip -20pt UWThPh-2004-18}
\prammonthinfo{\hskip -16.8pt hep-ph/0407178}
\prampagespace{\hskip 41pt}

\def\greaterthansquiggle{\raise.3ex\hbox{$>$\kern-.75em\lower1ex\hbox{$\sim$}}}
\def\lessthansquiggle{\raise.3ex\hbox{$<$\kern-.75em\lower1ex\hbox{$\sim$}}}
\newcommand{\beq}{\begin{equation}}
\newcommand{\eeq}{\end{equation}}
\newcommand{\beqa}{\begin{eqnarray}}
\newcommand{\eeqa}{\end{eqnarray}}
\newcommand{\ba}{\begin{array}}
\newcommand{\ea}{\end{array}}

\newcommand{\gsim}{\;\raisebox{-0.9ex}
           {$\textstyle\stackrel{\textstyle >}{\sim}$}\;}

\newcommand{\lsim}{\;\raisebox{-0.9ex}{$\textstyle\stackrel{\textstyle<}
           {\sim}$}\;}

\begin{document}
\mark{{Physics Possibilities at a Linear Collider}{Alfred Bartl and Stefan Hesselbach}}
\title{Physics Possibilities at a Linear Collider\footnote{Invited
plenary talk presented by A.~Bartl at the
``Workshop on High Energy Physics Phenomenology (WHEPP-8)'',
Indian Institute of Technology, Mumbai, January 5 -- 16, 2004.}}

\author{Alfred Bartl and Stefan Hesselbach}
\address{Institut f\"ur Theoretische Physik, Universit\"at Wien,
Boltzmanngasse 5, A-1090 Vienna, Austria}

\abstract{We review some recent studies about the parameter
determination of top quarks, $W$ bosons, Higgs bosons, supersymmetric
particles and in the ADD model of extra dimensions at a linear collider.
}

\maketitle
\section{Introduction}

Up to now the Standard Model (SM) has passed all accelerator-based
experimental tests.
It is able to reproduce all experimental data obtained at high energy
$e^+e^-$, $p \bar{p}$ and $e^{\pm} p$ colliders.
In particular, the precision data of LEP have verified the SM predictions
with very high accuracy, with the experimental errors being in the range
of about $0.1\,\% - 1\,\%$. On the theoretical side, most of the one-loop
corrections to the prominent observables have been calculated, in some
cases also the leading two-loop corrections are known. The theoretical
errors are also in the $0.1\,\% - 1\,\%$ range.

While the gauge sector of the SM is extremely well tested,
our theoretical ideas about electroweak symmetry breaking are still
not completely convincing.
In fact, clarifying the mechanism of electroweak symmetry breaking will be
the central problem which we have to solve with the next generation
of high energy colliders.
In the SM electroweak symmetry breaking
is achieved by the Higgs mechanism. The scalar Higgs boson, also predicted
by this mechanism, has not been found so far. The mass of the Higgs boson
is a free parameter of the theory. If the Higgs mechanism of the SM is
correct, then we know already that the mass of the Higgs boson is
constrained by the data of LEP: Its lower bound
is $m_H \gsim 114$ ~GeV by the direct searches, while from the fits to the
precision data an upper bound of $m_H \lsim 250$~GeV can be derived.
If the mass of the SM Higgs boson is indeed in this range,
then there is no doubt that it will be found at LHC.

The Higgs mechanism in its simplest form
like in the SM may cause quite
severe theoretical problems, namely the so-called ``hierarchy problem''
and the ``fine-tuning problem''. In order to cure these problems,
several new theoretical
ideas have been proposed, the most important ones are supersymmetry (SUSY),
compositeness, strong electroweak symmetry breaking, extra dimensions etc.
In many of these models the Higgs sector is more complicated than in the SM.
For example, in the Minimal
Supersymmetric Standard Model (MSSM) or in a more general Two-Higgs-Doublet
Model the Higgs sector contains two Higgs doublets.
The Higgs sector may also contain additional Higgs singlets or may
be even more complicated. In all of these cases we will probably see a
Higgs state at LHC, however, we may not be able to decide whether it is
the Higgs boson of the SM or it is a state of a more complicated
Higgs sector.

In some of the different theoretical models which have been proposed as
alternatives to the Higgs mechanism, electroweak symmetry breaking is
achieved without introducing an elementary
scalar Higgs particle. This is the case, for example, in technicolour
models or models of strong electroweak symmetry breaking, in models of
compositeness and others.
All these theoretical ideas have different phenomenological implications
which have to be tested in experiment.
As has been demonstrated in several recent workshops,
an $e^+e^-$ linear collider with a centre-of-mass energy in the range of
about $500$~GeV to $1$~TeV will enable us to distinguish between the
different mechanisms of electroweak symmetry breaking and will presumably
provide a unique answer to the questions about the origin of mass
\cite{acco,LC1,LC2,LC3}.

A new machine like an $e^+e^-$ linear collider must be able to improve
our knowledge in two ways: On the one hand it must allow us to discover
``new physics'', on the other hand it must also
provide a better understanding of ``known physics'' by more precise
measurements of the various SM parameters. In the recent workshops
\cite{LC1,LC2,LC3} it has
been demonstrated that an $e^+e^-$ linear collider will be a very good
discovery machine as well as an excellent precision instrument for physics
of and beyond the SM. At a linear collider we will be able to measure
the SM parameters like the masses of the top quark and the $W^{\pm}$ boson,
the running strong coupling constant $\alpha_s$ and many others with much
better precision than presently available. This machine will also allow us
to determine the parameters of the Higgs and SUSY sectors and other
extensions of the SM with very high accuracy.
As the signal cross sections decrease with increasing c.m.s.\ energy,
a high luminosity will
be required. It is expected that the integrated luminosity will reach a
value of approximately ${\cal L} = 500$~fb$^{-1}$ per year of running.
Also a high degree of beam polarisation will be
necessary \cite{Moortgat-Pick:2001kg}.
As has been shown, a degree of $80\,\%$ and $60\,\%$ for the $e^-$
and $e^+$ beam, respectively, can be achieved. Presently there are also
discussions about the possible benefits of transverse beam polarisations
\cite{power,RindaniLCWStalk}. Furthermore, it is expected
that the linear collider can be operated also in the $e^-e^-$, $e \gamma$
and $\gamma \gamma$ mode
(for reviews see \cite{eminuseminus}).
There is also the proposal to operate the
linear collider in the $e^+e^-$ mode at or near $\sqrt{s} \approx m_Z$,
which is also called the $GigaZ$ mode.

In this talk we will give a selective review of some of the recent studies
on the physics possibilities at a linear collider. We will select a few
illustrative examples in SM physics and in ``new physics'' to demonstrate
which new results we can expect to gain at such a new machine. For more
complete reviews we refer to \cite{acco,LC1,LC2,LC3}.

\section{Standard Model Physics}

We have at least two reasons why also in the future we have to perform
precision tests of the SM with increasing accuracy:
\begin{itemize}
\item We expect that the SM is only an effective theory valid at low
energies and we have to find the limits of its validity.
\item Usually any ``new physics'' reaction has a background of SM reactions,
which we must know with sufficient precision to extract the signal.
\end{itemize}

\begin{figure}[ht!]
\centerline{\epsfbox{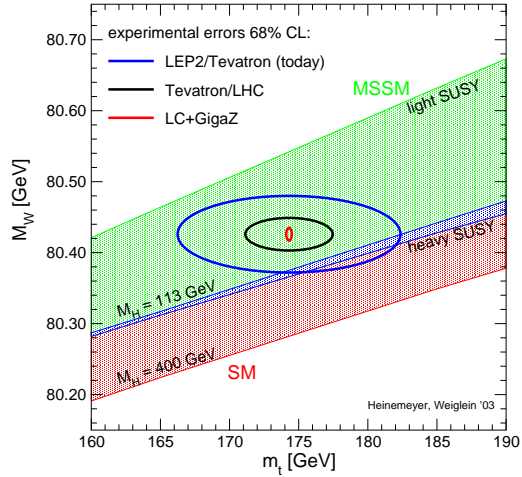}}
\caption{\label{fig:heiwei}
The present experimental accuracy for $m_W$ and $m_t$ 
after the experiments at LEP and Tevatron 
%(large ellipse)
(blue)
and the expected accuracies at Tevatron + LHC 
%(intermediate ellipse)
(black)
and LC + GigaZ 
%(small ellipse).
(red).
The 
%upper and lower
red and green 
bands show the predictions of SM and MSSM,
respectively, where the 
%small intermediate band
blue band 
denotes the overlap between the
predictions of SM and MSSM.
From \protect\cite{Heinemeyer:2003in}.}
\end{figure}

The present situation after the experiments at LEP, SLC and Tevatron can be
illustrated by an example from \cite{Heinemeyer:2003in} shown in
figure~\ref{fig:heiwei},
which is the result of a global analysis of the electroweak
precision data. This figure shows the theoretical relation
between the $W^{\pm}$ mass $m_W$ and the
top quark mass $m_t$ obtained in the SM 
%(lower band)
(red band)
and in MSSM 
%(upper band)
(green band)
together with the experimental error ellipses.
This theoretical relation between $m_W$ and $m_t$ is due to the radiative
corrections to the $W^{\pm}$ boson mass, where the loops involving the top
quark play a special role. The leading corrections depend quadratically on
$m_t$ and logarithmically on the Higgs boson mass $m_H$.
While in this calculation essentially all basic
electroweak parameters enter, $m_W$ depends very significantly on $m_t$
and on $m_H$.
The width of the SM band is mainly due to the variation of the Higgs boson
mass in the range $113~\textrm{GeV} \lsim m_H \lsim 400$~GeV. The MSSM band is
obtained by varying the SUSY parameters in the range allowed by the
experimental and theoretical constraints. There is a small overlap of
the SM and MSSM bands
%(small intermediate band)
(blue band)
for a light Higgs boson ($m_H = 113$~GeV) and a heavy SUSY
spectrum. As can be seen, the present experimental errors 
%(large ellipse)
(blue ellipse) 
do not allow to discriminate between the two models. The data of the linear
collider 
%(small ellipse)
(red ellipse) 
will presumably allow us to discriminate between the
SM and the MSSM or another extension of the SM.

In the calculation of the various observables at the $Z$ resonance the basic
electroweak parameters enter, which are the $Z$ boson mass $m_Z$,
the electromagnetic fine structure constant $\alpha$, the Fermi
coupling constant $G_F$, and the vector and axial-vector couplings $g^f_V$ and
$g^f_A$ of the $Z$ boson to the leptons and quarks. $\alpha$, $G_F$ and
$m_Z$ are known with very high accuracy. The coupling constants $g^f_V$ and
$g^f_A$ are determined by measurements at the $Z$ resonance, as there are
the $Z$ partial decay widths, forward-backward asymmetries,
$\tau$ polarisation and its forward-backward asymmetry at LEP, and the
left-right asymmetry and the left-right forward-backward asymmetry at
SLD \cite{lepewg}. The effective electroweak mixing angle
$\sin^2 \theta^f_{\rm eff}$ can then be determined via
$g^f_V/g^f_A = 1-4|q_f| \sin^2\theta^f_{\rm eff}$.
The weighted average of the experimental result for the
leptonic effective electroweak mixing angle is
$\sin^2 \theta^{lept}_{\rm eff} = 0.23150 \pm 0.00016$ \cite{lepewg}. In this
context it is interesting to note that by operating the linear collider
in the GigaZ mode the errors of all the $Z$ couplings could be
significantly reduced, leading to an error on the weak mixing angle of
approximately
$\delta \sin^2 \theta_{\rm eff} \approx \pm 1 \times 10^{-5}$ \cite{gigaZ}.

Note that in figure~\ref{fig:heiwei} for the top quark mass the value
$m_t = 174.3 \pm 5.1$~GeV has been used. This value has been changed
recently to $m_t = 178.0 \pm 4.3$~GeV \cite{mtop}. When using this new
value for $m_t$  the present error ellipse 
%(large ellipse)
(blue ellipse)
will be shifted
to the right by approximately $3.7$~GeV. On the one hand this improves the
agreement between the data and the SM predictions, giving
$m_H = 113^{+62}_{-42}$~GeV for the most probable value of the SM Higgs
mass. On the other hand
it also shows that, apart from finding the Higgs boson, it would be very
desirable to reduce the experimental errors of the masses of the
$W^{\pm}$ boson and the top quark. We will shortly review in the next
subsections which accuracy
for the top quark, $W^\pm$ boson and Higgs boson parameters
will be obtained at a linear collider.

\subsection{Top Quark Production}

Due to its high mass the top quark plays a special role among the SM
particles and it is often claimed that measuring all its parameters
very precisely may already open a window to new physics 
\cite{HiokiLCWStalk}. As
has been shown in \cite{Heinemeyer:2003ud}, for example, the precise
knowledge of
the top quark mass is necessary for the theoretical predictions of
electroweak precision observables and for the clarification of several
aspects of the Higgs sector and of the MSSM as well as for testing
grand unification. At the Tevatron and LHC
it is expected that the experimental error of the top quark mass can
be reduced to approximately $\delta m_t = 1$ -- $2$~GeV (see
e.g.~\cite{Beneke:2000hk}). At a linear collider an accuracy for the top
quark mass of about $\delta m_t \lesssim 100$~MeV is envisaged
\cite{LC1,LC2,LC3}.

The most precise determination of the top quark mass will be possible
by a threshold scan of $e^+ e^- \to t \bar{t}$ in the region
$340~\textrm{GeV} \lesssim\sqrt{s}\lesssim 380$~GeV. In order to
extract a precise value for $m_t$, also
theoretical work is necessary. As the top quark
width $\Gamma_t \approx 1.5$~GeV is much larger than
$\Lambda_{\mathrm{QCD}}$, no toponium bound states are formed,
however, their traces may be present in the threshold region. For the
theoretical analysis it is necessary to perform a double-expansion of
the production cross section $\sigma_{t\bar{t}}$ in the strong
coupling $\alpha_s$ and the velocity $v$ of the top quark,
$v = \sqrt{1 - (4 m_t^2)/s} \ll 1$. The terms of this expansion
up to NNLO have been calculated \cite{Hoang}.
There are large logarithms which have been resummed. The error of
$m_t$ is correlated with that of $\alpha_s$, however, a simultaneous
precision measurement of $m_t$ and $\alpha_s$ by a threshold scan will
be possible. The analyses of 
\cite{LC1,LC2,LC3,Hoang,Martinez:2002st,BrandenburgEcfaTalk} 
indicate that the
expected errors are $\delta m_t^{\mathrm{exp+theo}} \approx 100$~MeV,
$\delta\alpha_s(M_Z)^{\mathrm{exp}} \sim 0.001$, and for the total width 
$\delta\Gamma_t^{\mathrm{exp}} \sim 50$~MeV.

Furthermore, by measuring various decay
distributions we will obtain information about the top quark
couplings. For example, the single decay-lepton angular distributions
and angular asymmetries are sensitive to the CP-violating top quark
couplings \cite{rindani,GrzadkowskiHioki}.

\subsection{$W^{\pm}$ Boson Production}

Operating the linear collider near the kinematical threshold of the
reaction $e^+ e^- \to W^+ W^-$ will provide a measurement of the
$W^\pm$ boson mass with an error of $\delta_{m_W} \approx \pm 6$~MeV.
Analysing cross sections with polarised beams and various decay
distributions will allow us to determine the triple gauge boson
couplings with high accuracy.

The most general Lagrangian describing anomalous triple gauge boson
couplings is given in \cite{Hagiwara:1986vm} and contains 28 real
parameters, which can be parameterised by 7 complex coupling parameters
$g_1^V$, $\kappa_V$, $\lambda_V$, $g_5^V$, $g_4^V$,
$\tilde{\kappa}_V$ and $\tilde{\lambda}_V$, where $V=\gamma,Z$.
The first four couplings are CP conserving and the last three CP
violating, whereas the real (imaginary) parts of all couplings are 
CP$\tilde{\textrm{T}}$ conserving (violating), where
$\tilde{\textrm{T}}$ denotes the ``naive'' time reversal transformation.
In the SM all these couplings are zero except
$g_1^\gamma = g_1^Z = \kappa_\gamma = \kappa_Z = 1$.
$g_1^\gamma$ determines the charge of the $W$ and $\kappa_\gamma$,
the ``anomalous magnetic moment'', and
$\lambda_\gamma$ are related to the magnetic dipole moment and the electric
quadrupole moment of the $W$, whereas
$\tilde{\kappa}_\gamma$ and $\tilde{\lambda}_\gamma$ are related to the
electric dipole moment and the magnetic quadrupole moment.

In \cite{Diehl:2003qz} the prospects of measuring the triple gauge
couplings at a linear collider with transverse beam polarisation in
the process $e^+e^- \to W^+ W^-$ are studied.
The expected errors for the CP and CP$\tilde{\textrm{T}}$ conserving 
couplings are given in table~\ref{tab:errTGC}.
The sensitivity in the other three (CP, CP$\tilde{\textrm{T}}$) symmetry
classes is comparable and there is no statistical correlation between
the classes. Within a symmetry class there are large statistical
correlations.
Without beam polarisation 27 coupling parameters appear to be
measurable, longitudinal $e^-$ beam polarisation $P_{e^-}$ improves
the sensitivity by a factor $\sim 2$, additional longitudinal
$e^+$ beam polarisation $P_{e^+}$ by a factor $3$ -- $4$.
With transverse beam polarisation ($P_{e^-}^t,P_{e^+}^t$)
all couplings may be accessible at a linear collider and there is a
factor $2$ -- $4$ improvement in comparison to unpolarised beams.
The couplings $\kappa_\gamma$ and $\lambda_\gamma$ have also been
analysed at a photon collider \cite{Bozovic-Jelisavcic:2002ta}.
With help of the process $e^- \gamma \to W^- \nu$
it is possible to improve the sensitivity on $\lambda_\gamma$ by a
factor $1.5$ in comparison to the process $e^+ e^- \to W^+ W^-$ with
both beams polarised, whereas the sensitivity on $\kappa_\gamma$ is of
the same order of magnitude.

\begin{table}[ht]
\caption{\label{tab:errTGC} Expected errors in units of $10^{-3}$ on
the CP and CP$\tilde{\textrm{T}}$ conserving
couplings in the presence of all anomalous
couplings at \mbox{$\sqrt{s}=$ 500 GeV}, with unpolarised beams and
with different beam polarisations.
From \protect\cite{Diehl:2003qz}.}
\begin{center}
\begin{tabular} {l|cccccccc}
 & Re$\,\Delta g_1^{\gamma}$ & Re$\,\Delta g_1^Z$ & Re$\,\Delta
\kappa_{\gamma}$ & Re$\,\Delta \kappa_Z$ & Re$\,\lambda_{\gamma}$ &
Re$\,\lambda_Z$ & Re$\,g_5^{\gamma}$ & Re$\,g_5^Z$ \rule{0mm}{9pt}\\
\hline
 Unpolarised beams & 6.5 & 5.2 & 1.3 & 1.4 & 2.3 & 1.8 & 4.4 & 3.3\\
$(P_{e^-},P_{e^+})=(\mp 80\,\%,0)$ & 3.2 & 2.6 & 0.61 & 0.58 & 1.1 & 0.86 &
2.2 & 1.7\\
$(P_{e^-},P_{e^+})=(\mp 80\,\%,\pm 60\,\%)$ & 1.9 & 1.6 & 0.40 & 0.36 & 0.62 &
0.50 & 1.4 & 1.1\\
$(P_{e^-}^t,P_{e^+}^t)=(80\,\%,60\,\%)$ & 2.8 & 2.4 & 0.69 & 0.82 & 0.69 & 
0.55 & 2.5 & 1.9
\end{tabular}
\normalsize
\end{center}
\end{table}

\subsection{Higgs Boson Production}

\begin{figure}[ht!]
\centerline{\epsfbox{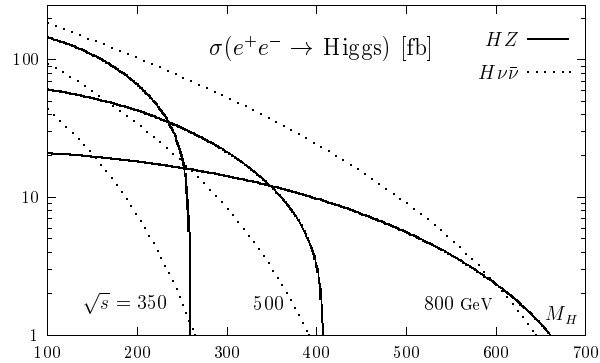}}
\caption{\label{fig:SMHiggs}
The Higgsstrahlung and $WW$ fusion production cross sections versus
the Higgs mass for $\sqrt{s} = 300$~GeV, $500$~GeV and $800$~GeV.
From \protect\cite{LC2,Djouadi:2002pw}.}
\end{figure}

The search for the Higgs boson will have a very high priority at LHC
as well as at the linear collider \cite{LC1,LC2,LC3,Djouadi:2002pw}.
The production of a SM Higgs boson
in $e^+ e^-$ annihilation can proceed via ``Higgsstrahlung''
$e^+ e^- \to Z H$, $WW$ fusion $e^+ e^- \to \nu_e \bar{\nu}_e H$, and
$ZZ$ fusion $e^+ e^- \to e^+ e^- H$.
The relative importance of Higgsstrahlung and $WW$ fusion is shown in
figure~\ref{fig:SMHiggs} (from \cite{LC2,Djouadi:2002pw}).
At $\sqrt{s} = 500$~GeV the Higgsstrahlung process dominates for
$m_H \gtrsim 160$~GeV, whereas for $m_H \lesssim 160$~GeV the $WW$
fusion process gives the largest contribution. The higher $\sqrt{s}$
the more important is the $WW$ fusion process. For
$\sqrt{s} \gtrsim 800$~GeV the $ZZ$ fusion process can contribute
about $\gtrsim 10\,\%$ of the total production rate.
As can also be seen in figure~\ref{fig:SMHiggs}, if
$m_H \lesssim 250$~GeV as suggested by the electroweak precision data,
an optimal choice for the c.m.s.\ energy is $\sqrt{s} \approx 350$ --
$500$~GeV.

\begin{figure}[ht!]
\centerline{\epsfbox{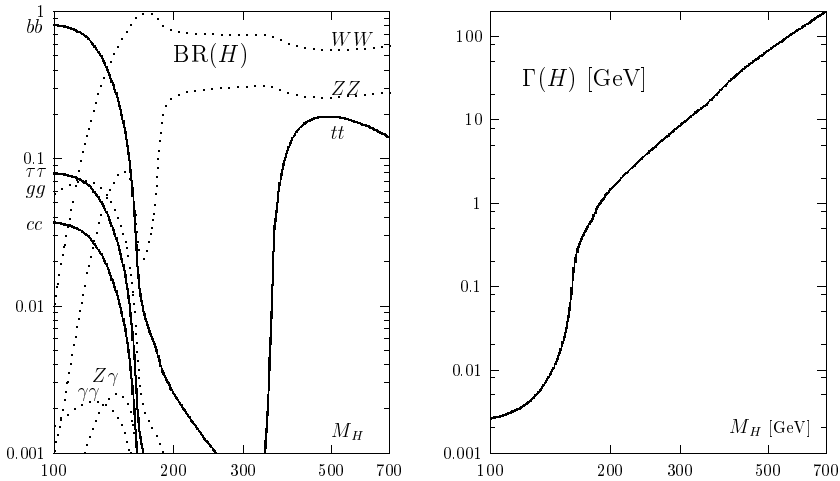}}
\caption{\label{fig:SMHiggsBR}
(a) Branching ratios and (b) the total decay width
of the SM Higgs boson as a function of its mass.
From \protect\cite{LC2,Djouadi:2002pw}.}
\end{figure}

In order to distinguish a SM Higgs boson from a state of a more
complicated Higgs sector it will be necessary to measure its decay
branching ratios with good precision. The couplings of the SM Higgs to
the fermions and vector bosons are proportional to their
masses. Hence, if the Higgs boson mass is fixed, then also all decay
widths are known. In figure~\ref{fig:SMHiggsBR} from 
\cite{LC2,Djouadi:2002pw} we show
the most important branching ratios of the SM Higgs and the total
width as a function of its mass. As can be seen, beyond the threshold
of the decay $H \to W^+ W^-$ the width rises rather strongly with the
mass. For $m_H < 150$~GeV the decay $H \to b \bar{b}$ dominates,
whereas the branching ratios for $H \to \tau^+ \tau^-$, $H \to c \bar{c}$
and $H \to g g$ are of the order of a few percent. It is expected that
for a light Higgs boson the coupling to the bottom quark can be
determined with an error of about $2\,\%$ and those to the charm quark
and the $\tau$ lepton with about $10\,\%$. The coupling of the SM
Higgs boson to the top quark is particularly interesting. For
$m_H < 2 m_t$ a possible way to determine this coupling is given by
measuring the associated production process $e^+ e^- \to t \bar{t} H$. 
For measuring this coupling with an error of about $10\,\%$ we will
need higher energy, $\sqrt{s} \gtrsim 800$~GeV, and higher luminosity,
${\cal L} \approx 1000~\textrm{fb}^{-1}$ per one year of running.

A further test for the scalar potential of the SM Higgs boson will be
provided by measurements of its self-couplings. The 3-Higgs coupling
is given by $\lambda_3 = m_H^2/(2 v)$ and the 4-Higgs coupling by
$\lambda_4 = m_H^2/(2 v^2)$, where $v = 246$~GeV is the vacuum
expectation value of the Higgs field after spontaneous electroweak
symmetry breaking. The coupling $\lambda_3$ can be measured in the
``double-Higgsstrahlung'' process $e^+ e^- \to Z H H$.
The analyses of \cite{Higgsselfcouplings,Djouadi:1999gv} have shown
that $\lambda_3$ can be measured with an error of approximately
$20\,\%$ in the Higgs mass range
$120~\textrm{GeV} < m_H < 140~\textrm{GeV}$ with an integrated
luminosity of $1000~\textrm{fb}^{-1}$. The 4-Higgs coupling
$\lambda_4$ can in principle be measured in the
``triple-Higgsstrahlung'' process, which however is suppressed
\cite{Djouadi:1999gv}.

It is also necessary to verify that the Higgs boson is a spin $0$ CP
even particle. At the linear collider this can be done in several
ways:
(i) By measuring the angular dependence of $e^+ e^- \to Z H$, which
must be $d\sigma/d\cos\theta \propto \sin^2\theta$ for a $0^+$
particle.
(ii) By measuring the $\sqrt{s}$ dependence of $e^+ e^- \to Z H$ near
threshold.
(iii) If $m_H < 2 m_Z$ by measuring the angular dependence of the
decay products in the decay $H \to Z Z^*$ and by measuring the
invariant mass distribution of the virtual $Z$.

\section{Beyond the standard model}

\subsection{Extended Higgs sectors}

In models beyond the SM additional Higgs doublet and
singlet fields can occur.
In the Two-Higgs-Doublet Models, the simplest of these extended
models, the Higgs sector consists of two doublet fields, which
generate five physical Higgs particles:
the CP-even $h$ and $H$, the CP-odd $A$ and the charged $H^\pm$.
In particular the Higgs sector of the Minimal Supersymmetric Standard
Model (MSSM) contains two doublets, which are necessary to break the
electroweak symmetry.

Supersymmetry leads to several relations among the parameters of the
MSSM Higgs sector resulting at tree-level
in only two independent parameters and mass relations like
$m_h \le m_Z, m_A \le m_H$; $m_{H^\pm} \ge m_W$; 
$m_{H^\pm}^2 = m_A^2 + m_W^2$; $m_h^2 + m_H^2 = m_A^2 + m_Z^2$.
Radiative corrections, however, break some of these correlations and
especially $m_h$ is pushed upwards by several ten GeV.
A recent analysis, including full one-loop corrections and the two-loop
corrections controlled by $\alpha_s$ and the Yukawa couplings of the
third generation fermions and using the latest value of the top quark
mass \cite{mtop}, gives a conservative upper bound 
$m_h \lesssim 152$~GeV \cite{Allanach:2004rh}.

\begin{figure}[ht!]
\centerline{\epsfbox{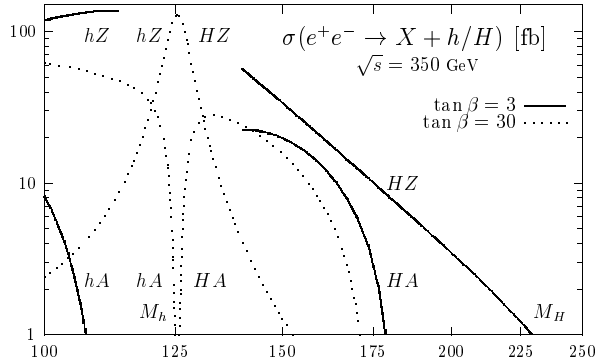}}
\caption{\label{fig:MSSMHiggs}
Production cross sections of the neutral Higgs bosons in the MSSM at
$\sqrt{s}=350$~GeV in the Higgsstrahlung and pair production
processes for $\tan\beta = 3$ and $30$.
From \protect\cite{LC2,Djouadi:2002pw}.}
\end{figure}

In the two-Higgs-doublet extensions of the SM the CP-even Higgs bosons
$h$ and $H$ can be produced by the associated production process
$e^+ e^- \to A + h,H$ in addition to Higgsstrahlung and $WW$ fusion.
The CP-odd Higgs boson $A$ cannot be produced in Higgsstrahlung and
fusion processes at leading order. The charged Higgs bosons $H^\pm$
can be directly pair produced, $e^+ e^- \to H^+ H^-$.
Figure~\ref{fig:MSSMHiggs} from
\cite{LC2,Djouadi:2002pw} shows representative examples
of $e^+ e^- \to Z + h,H$ and $e^+ e^- \to A + h,H$ 
production cross sections in the MSSM as a function of the respective
Higgs masses.

In the CP-violating MSSM with complex parameters the CP invariance of
the Higgs potential at tree-level is explicitely broken by loop
effects of the Yukawa interactions of the third generation squarks
\cite{CPHiggs}. This results in a mixing of the neutral Higgs states $h$,
$H$ and $A$ into three mass eigenstates $H_1$, $H_2$ and $H_3$.
This mixing can considerably change the mass spectrum and couplings of
the neutral Higgs bosons and has to be taken into account when
analysing the complex MSSM.

\subsection{Supersymmetry}

One of the main goals of a future $e^+e^-$ linear collider will be the
search for supersymmetric (SUSY) particles \cite{acco,LC1,LC2,LC3}. 
The neutralinos, the supersymmetric partners of the neutral gauge and
Higgs bosons, will be particularly interesting because they are
relatively light. The lightest neutralino $\tilde{\chi}^0_1$ is
expected to be the lightest SUSY particle (LSP), which is stable if
$R$-parity is conserved and the second lightest neutralino
$\tilde{\chi}^0_2$ will presumably be among the lightest visible SUSY
particles. 
Therefore, the study of production and
decay of the neutralinos $\tilde{\chi}^0_i$, $i=1,\ldots,4\,$,
\cite{acco,LC1,LC2,LC3,Bartl:hp}
and a precise determination of the underlying supersymmetric
parameters $M_1$, $M_2$, $\mu$ and $\tan\beta$
including the phases $\phi_{M_1}$ and $\phi_\mu$ of
$M_1$ and $\mu$ will play an important role at future linear colliders.
In \cite{NeuChaParDet} methods to determine these parameters based on
neutralino and chargino mass and cross section measurements have been
presented.

\begin{figure}[ht!]
\centerline{\epsfbox{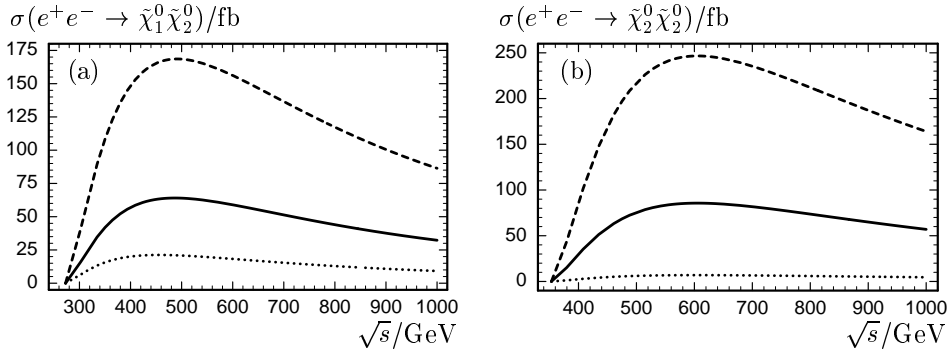}}
\caption{\label{fig:neutprod}
Cross sections for
(a) $e^+e^- \to \tilde{\chi}^0_1 \tilde{\chi}^0_2$ and 
(b) $e^+e^- \to \tilde{\chi}^0_2 \tilde{\chi}^0_2$
in the scenario SPS1a \protect\cite{sps}
for unpolarised beams (solid) and beam polarisations
$P_{e^-} = -0.8$, $P_{e^+} = +0.6$ (dashed),
$P_{e^-} = +0.8$, $P_{e^+} = -0.6$ (dotted).}
\end{figure}

The production of neutralinos 
$e^+e^- \to \tilde{\chi}^0_i \tilde{\chi}^0_j$, $i,j=1,\ldots4$, 
at a linear collider with polarised beams
with subsequent leptonic tree-body decays 
$\tilde{\chi}^0_i \to \tilde{\chi}^0_k \ell^+ \ell^-$
and $\tilde{\chi}^0_j \to \tilde{\chi}^0_l \ell^+ \ell^-$, 
is analysed in \cite{Moortgat-Pick:1999di}.
Since observables like decay angular distributions and T-odd triple
product correlations depend on the polarisation of the decaying
neutralinos \cite{spincorr}
the full spin correlations between production and decay
are included.
In \cite{Moortgat-Pick:1999di} the complete analytical formulae for
longitudinal polarised beams are given, including complex couplings to
allow the study of CP violating phenomena.
In figure~\ref{fig:neutprod} the cross sections
$\sigma(e^+e^- \to \tilde{\chi}^0_1 \tilde{\chi}^0_2)$ and
$\sigma(e^+e^- \to \tilde{\chi}^0_2 \tilde{\chi}^0_2)$
in the scenario SPS1a \cite{sps} are shown as a
function of the $e^+e^-$ centre of mass energy $\sqrt{s}$ for
several beam polarisations.

The pair production of light sleptons
$e^+ e^- \to \tilde{\ell}^+ \tilde{\ell}^-$,
$\tilde{\ell} = \tilde{e}_R, \tilde{\mu}_R, \tilde{\tau}_1$,
allows to determine the masses of the sleptons with high accuracy
from the energy distributions of the final state particles in the
slepton decays $\tilde{\ell} \to \ell \, \tilde{\chi}^0_1$.
The masses of $\tilde{e}_R$ and $\tilde{\mu}_R$ can be determined
from the energy spectrum of the final electron and muon, respectively
and $m_{\tilde{\tau}_1}$ from the energy spectra $E_\rho$ and
$E_{3\pi}$ from the $\tau$ decays $\tau \to \rho \, \nu_\tau$ and
$\tau \to 3\pi \, \nu_\tau$. 
The study in \cite{Martyn:2004ew} for the 
TESLA linear collider \cite{LC2}
in the scenario SPS1a \cite{sps} expects the
accuracies $m_{\tilde{e}_R} = (142.99 \pm 0.08)$~GeV,
$m_{\tilde{\mu}_R} = (143.15 \pm 0.17)$~GeV and
$m_{\tilde{\tau}_1} = (133.2 \pm 0.3)$~GeV
for $\sqrt{s} = 400$~GeV, 
beam polarisations $P_{e^-} = +0.8$, $P_{e^+} = -0.6$ and
an integrated luminosity ${\cal L} = 200~\textrm{fb}^{-1}$.

\begin{figure}[ht!]
\centerline{\epsfbox{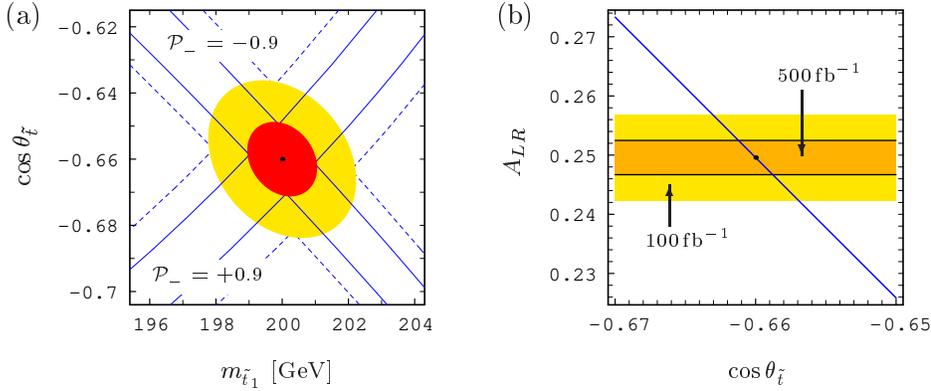}}
\caption{\label{fig:stopmix}
(a) Error bands and 68\,\% CL error ellipse for determining the
$\tilde{t}_1$ mass $m_{\tilde{t}_1}$
and mixing angle $\cos\theta_{\tilde{t}}$ from cross section measurements;
the dashed lines are for ${\cal L}=100~\mathrm{fb}^{-1}$ and
the full lines for ${\cal L}=500~\mathrm{fb}^{-1}$.
(b) Error bands for the determination of $\cos\theta_{\tilde{t}}$
from the left-right asymmetry $A_{LR}$.
In both plots $m_{\tilde{t}_1}=200$ GeV,
$\cos\theta_{\tilde{t}}=-0.66$, 
$\sqrt{s}=500$ GeV,
$P_{e^-}=\pm 0.9$, $P_{e^+}=0$.
From \protect\cite{Bartl:2000kw}.}
\end{figure}

Whereas the parameters $M_1$, $M_2$ and $\mu$
including the phases $\phi_{M_1}$ and $\phi_\mu$
can
be determined by measurements of CP-even \cite{NeuChaParDet} and
CP-odd \cite{Choi:1999cc,ATtwobody,Aguilar-Saavedra:2004dz,Bartl:2004jj}
observables in the neutralino and chargino sector, it is more
difficult to measure the trilinear couplings $A_f$ in the
sfermion sector.
Cross section measurements of sfermion production processes allow the
determination of the sfermion masses and mixing angles (see
figure~\ref{fig:stopmix}) which in turn allow the determination of the
parameters $A_f$ \cite{Bartl:2000kw}.

In \cite{staupapers,squarkpapers} production and decays of the third
generation sfermions in the MSSM with complex parameters $A_\tau$,
$A_t$, $A_b$, $\mu$ and $M_1$ have been analysed.
In a large region of the MSSM parameter space the branching
ratios of $\tilde{\tau}_{1,2}$, $\tilde{\nu}_\tau$, $\tilde{t}_{1,2}$
and $\tilde{b}_{1,2}$ show a strong phase dependence.
This could have an important impact on the search
for third generation sfermions at a future linear collider and
on the determination of the supersymmetric parameters.

In \cite{staupapers} the effects of the CP phases of $A_\tau$, $\mu$
and $M_1$ on production and decay of $\tilde{\tau}_{1,2}$ and
$\tilde{\nu}_\tau$ have been studied.
The branching ratios of fermionic decays of $\tilde{\tau}_{1}$ and
$\tilde{\nu}_\tau$ show a significant phase dependence for $\tan\beta
\lesssim 10$ whereas it becomes less pronounced for $\tan\beta > 10$.
The branching ratios of the $\tilde{\tau}_{2}$ into Higgs bosons
depend very sensitively on the phases for $\tan\beta \gtrsim 10$.

\begin{figure}[ht!]
\centerline{\epsfbox{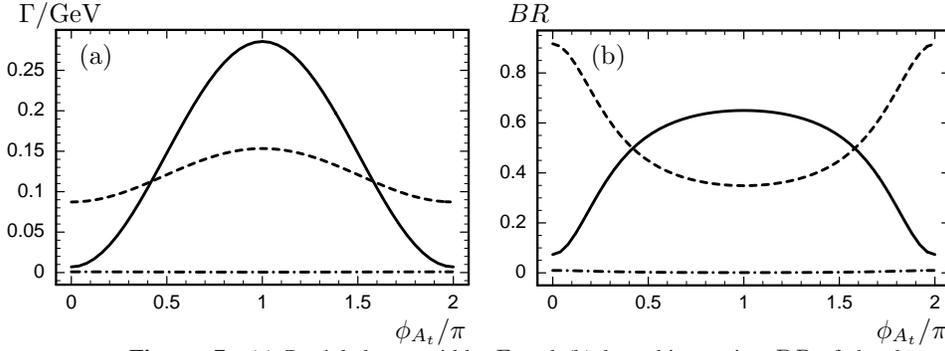}}
\caption{\label{fig:st1decays}
(a) Partial decay widths $\Gamma$
and (b) branching ratios $BR$ of the decays
$\tilde{t}_1 \to \tilde{\chi}^+_1 b$ (solid),
$\tilde{t}_1 \to \tilde{\chi}^0_1 t$ (dashed) and
$\tilde{t}_1 \to W^+ \tilde{b}_1$ (dashdotted)
for $\tan\beta = 6$, $M_2=300$~GeV, 
$M_1/M_2 = 5/3 \, \tan^2\theta_W$, $|\mu|=350$~GeV,
$|A_b|=|A_t|=800$~GeV,
$\varphi_\mu=\pi$, $\varphi_{M_1}=\varphi_{A_b}=0$,
$m_{\tilde{t}_1}=350$~GeV, $m_{\tilde{t}_2}=700$~GeV,
$m_{\tilde{b}_1}=170$~GeV, $M_{\tilde{Q}}>M_{\tilde{U}}$
and $m_{H^\pm}=900$~GeV.
From \protect\cite{squarkpapers}.}
\end{figure}

In \cite{squarkpapers} the impact of the CP phases of $A_t$, $A_b$,
$\mu$ and $M_1$ on the decays of $\tilde{t}_{1,2}$ and
$\tilde{b}_{1,2}$ has been analysed.
The branching ratios of the $\tilde{t}_{1,2}$ show a pronounced phase
dependence in a large region of the MSSM parameter space
(figure~\ref{fig:st1decays}).
In the case of $\tilde{b}_i$ decays there can be an appreciable
$\varphi_{A_b}$ dependence, if $\tan\beta$ is large and the decays
into Higgs bosons are allowed.

Further the expected accuracy in determining the supersymmetric
parameters has been estimated by a global fit of measured masses, branching
ratios and production cross sections. $A_\tau$, $A_t$ and $A_b$
can be expected to be measured with 10\,\%, 2 -- 3\,\% and 50\,\%
accuracy, respectively,
$\tan\beta$ with 1\,\% (2\,\%) accuracy in case
of small (large) $\tan\beta$ and the other parameters with 
approximately 1\,\% accuracy.

\begin{figure}[ht!]
\centerline{\epsfbox{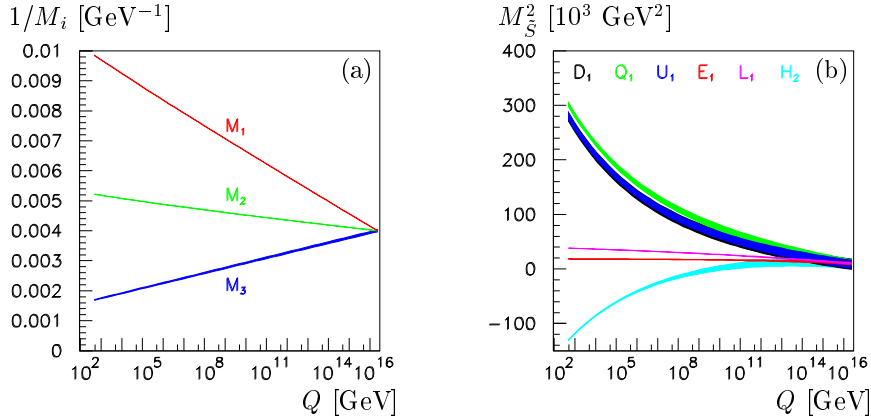}}
\caption{\label{fig:RGErunning}
Evolution of (a) the gaugino mass parameters and (b) the first
generation sfermion mass parameters from low to high scales $Q$ in the
SPS1a scenario \protect\cite{sps} for a coherent combination of LHC
and linear collider analyses.
From \protect\cite{Allanach:2004ud}.}
\end{figure}

The fundamental parameters at the GUT or unification scale in specific
SUSY breaking models can be reconstructed by evolution of the
measured parameters at the electroweak scale to the high scale with
help of the respective renormalization group equations (RGE).
In \cite{Blair:2002pg} the evolution in minimal supergravity (mSUGRA),
in left-right supergravity, in gauge mediated SUSY breaking and in
superstring induced SUSY breaking models is analysed and the RGE are
given.
In figure~\ref{fig:RGErunning} from \cite{Allanach:2004ud} the
evolution of the gaugino mass parameters and of the first generation
sfermion mass parameters is shown in the mSUGRA scenario SPS1a
\cite{sps}, where the experimental errors at the electroweak scale 
have been fixed by a coherent combination of the analyses
for the LHC and for a linear collider. This allows to determine the mSUGRA
parameters with the following accuracies:
$m_{1/2} = (250.0 \pm 0.2)$~GeV, $m_0 = (100.0 \pm 0.2)$~GeV,
$A_0 = (-100 \pm 14)$~GeV and $\tan\beta = 10.0 \pm 0.4$.

The phases $\phi_{M_1}$ and $\phi_\mu$ of $M_1$ and $\mu$
cause CP-violating effects
already at tree-level. A useful tool to study these CP-violating
effects are T-odd observables, based on triple products
of momenta or spin vectors of the particles involved
\cite{Choi:1999cc,tripleproducts}.
In \cite{Bartl:2004jj} a T-odd asymmetry
\begin{displaymath}
A_T = \frac{\int\mathrm{sign}\{
 O_T\}
 |T|^2d\mbox{lips}}{{\int}|T|^2d\mbox{lips}}
\end{displaymath}
is defined
for neutralino production and subsequent leptonic three-body decay
with help of the triple product
$O_T=\vec{p}_{\ell^+}\cdot (\vec{p}_{\ell^-}\times\vec{p}_{e^-})$
of the initial electron momentum $\vec{p}_{e^-}$ and
the two final lepton momenta $\vec{p}_{\ell^+}$ and $\vec{p}_{\ell^-}$.
Here ${\int}|T|^2d\mbox{lips}$ is proportional to
the cross section
$\sigma(e^+e^- \to \tilde{\chi}^0_i\tilde{\chi}^0_j
\to \tilde{\chi}^0_i\tilde{\chi}^0_k \ell^+ \ell^-)$.
The asymmetry $A_T$ can be directly measured in the experiment without
reconstruction
of the momentum of the decaying neutralino or further final-state analyses.
Analogous T-odd asymmetries in neutralino production
and subsequent two-body decays have been studied
in \cite{ATtwobody}.

\begin{figure}[ht!]
\centerline{\epsfbox{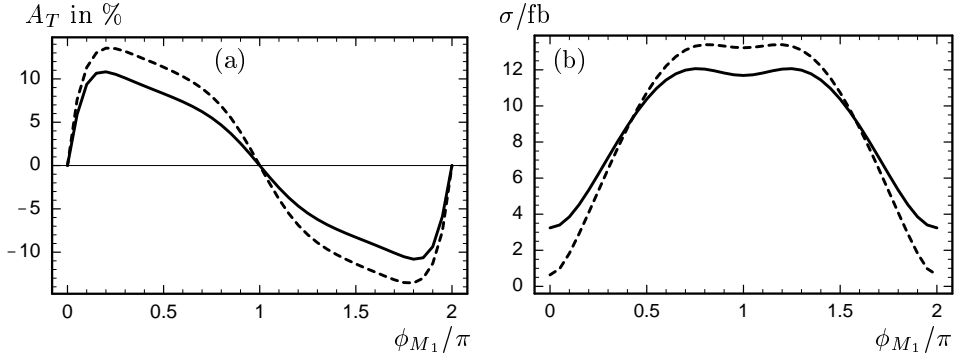}}
\caption{\label{fig:At12}
(a) CP asymmetry $A_T$ and (b) cross section
$\sigma(e^+e^- \to \tilde{\chi}^0_1\tilde{\chi}^0_2 \to$
$\tilde{\chi}^0_1\tilde{\chi}^0_1 \ell^+ \ell^-)$,
summed over $\ell=e,\mu\,$,
for $|M_1|=150$~GeV, $M_2=300$~GeV, $|\mu|=200$~GeV, $\tan\beta=10$,
$m_{\tilde{\ell}_L}=267.6$~GeV, $m_{\tilde{\ell}_R}=224.4$~GeV
and $\phi_{\mu}=0$ at a linear collider with beam polarisations
$P_{e^-}=-0.8$, $P_{e^+}=+0.6$ and
$\sqrt{s}=500$~GeV (solid), $\sqrt{s}=350$~GeV (dashed).
From \protect\cite{Bartl:2004jj}.}
\end{figure}

In figure \ref{fig:At12} the T-odd asymmetry $A_T$ and the
corresponding cross section at a linear
collider with polarised beams is shown in a representative scenario of
the unconstrained MSSM for the production 
$e^+e^-\to\tilde{\chi}^0_1\tilde{\chi}^0_2$
and the subsequent decay
$\tilde{\chi}^0_2\to \tilde{\chi}^0_1\ell^+\ell^-$.
For a centre of mass energy of 500~GeV (350~GeV) the asymmetry reaches
values $|A_T| = 11\,\%$ ($13.5\,\%$) for $\phi_{M_1} = 0.2\pi$ and
$1.8\pi$.
A Monte Carlo study of $A_T$ including initial state radiation,
beamstrahlung, SM backgrounds and detector effects
has been given in \cite{Aguilar-Saavedra:2004dz}.
It has been found that asymmetries $A_T \sim 10\,\%$ are detectable
after few years of running of a linear collider.

In the general SUSY Lagrangian
R-parity can be violated by bilinear and trilinear couplings in the
superpotential, which has two important
consequences. On the one hand lepton number violating couplings
contribute to the Majorana neutrino mass matrix
(for a review see \cite{Mukhopadhyaya:2003te}),
on the other hand the
LSP is not stable and hence not necessarily a neutral particle.
In sizeable regions of mSUGRA, GMSB or AMSB parameter space a charged
slepton can be the LSP.
Therefore, in \cite{Hirsch:2002ys,Bartl:2003uq}
the decay properties at colliders of
such charged slepton LSPs are studied. This can be used to obtain
information about the relative size of the bilinear and trilinear
couplings and to decide which of them give the dominant contribution to the
neutrino mass matrix.

In models, where neutrino oscillation properties are governed by
bilinear couplings only \cite{Hirsch:2002ys},
the slepton decay lengths are very different,
$L(\tilde{\tau}_1) \ll L(\tilde{\mu}_1) \ll L(\tilde{e}_1)$,
especially 
$L(\tilde{\tau}_1)/L(\tilde{\mu}_1) \sim m_\mu^2/m_\tau^2$.
Furthermore ratios of branching ratios are strongly correlated with
ratios of the bilinear couplings,
hadronic final states are never visible and
$BR(\tilde{e}_1 \to e \sum \nu_i) \approx 1$.

In models with trilinear couplings contributing to the neutrino masses
\cite{Bartl:2003uq} the decay lengths of the charged slepton LSPs
depend on the absolute values of the trilinear couplings and hence are
strongly correlated with neutrino observables.
Observing a finite decay length for $\tilde{e}_1$ or $\tilde{\mu}_1$
would imply that the corresponding trilinear couplings cannot
contribute significantly to the neutrino masses. 
Contrarily, the decay length of $\tilde{\tau}_1$ in models, where the
corresponding couplings determine the neutrino masses and where the
solar neutrino mass squared difference is in the correct order of
magnitude $5.1 < \Delta m_\odot^2/(10^{-5}~\mathrm{eV}^2) < 19$, is
just at the borderline of experimental accessibility.
Furthermore also hadronic final states may have visible branching
ratios and in models with
only trilinear couplings contributing to the neutrino masses all right
slepton LSPs obey $BR(\tilde{\ell}_1 \to (e,\mu,\tau) \sum \nu_i) < 0.5$.

\subsection{Extra dimensions}

A solution to the hierarchy problem can in principle be obtained by
formulating gravity in $4 + \delta$ dimensions, where
$\delta = 1,2,3,\ldots$ are the so-called ``extra'' dimensions, which
are assumed to be compactified  with a radius $R$.
In the model of \cite{add} it is assumed that SM physics is restricted
to the 4-dimensional brane, whereas gravity acts in the $4 + \delta$
dimensional bulk.
In 4-dimensional space-time the Planck mass is 
$M_{\mathrm{Pl}} = 1.2 \cdot 10^{19}$~GeV. In the 
$(4 + \delta)$-dimensional space the corresponding Planck mass $M_D$
is given by $M_D^{2+\delta} = M_{\mathrm{Pl}}^2/R^\delta$.
Assuming further that the compactification radius $R$ is many orders
of magnitude larger than the Planck length,
$R \gg M_{\mathrm{Pl}}^{-1}$, $R$ and $\delta$ may be adjusted such
that $M_D \approx {\cal O}(1~\textrm{TeV})$. In this way the Planck
scale is close to the electroweak scale and there is no hierarchy
problem \cite{add,ant}.

As a consequence of the compactification Kaluza-Klein towers of the
gravitons can be excited. This leads to two possible signatures at an
$e^+ e^-$ linear collider. The first one is 
$e^+ e^- \to \gamma/Z + G_n$ where $G_n$ means the graviton and its
Kaluza-Klein excitations, which appear as missing energy in the
detector. The second signature is due to graviton exchange in
$e^+ e^- \to f \bar{f}$, which leads to a modification of cross
sections and asymmetries compared to the SM prediction.

The cross section for $e^+ e^- \to \gamma/Z + G$ has been calculated
in \cite{Giudice:1998ck}. The main background to this process is
$e^+ e^- \to \nu \bar{\nu} \gamma$, which strongly depends on the
$e^-$ beam polarisation.
Table~\ref{tab:extradim} from \cite{LC2} shows the results on the
sensitivity in $M_D$ for various values of $\delta$.
Further aspects of ``extra dimensions'' physics can be found in
\cite{extraDthisproc,extraDfurther}.

\begin{table}[hbt]
\caption{\label{tab:extradim}
Sensitivity at $95\,\%$ CL in mass scale $M_D$ in TeV
for direct graviton production in $e^+ e^- \to \gamma G$ for
various values of $\delta$ taking a
$0.3\,\%$ normalisation error.
From \protect\cite{LC2}.}
\begin{tabular}{l|ccccc}
$\delta$                       &   2  &  3  &  4  &  5  &  6  \\ \hline
$M_D$ for $(P_{e^-},P_{e^+})=(0,0)$        &  5.9 & 4.4 & 3.5 & 2.9 & 2.5 \\
$M_D$ for $(P_{e^-},P_{e^+})=(80\,\%,0)$   &  8.3 & 5.8 & 4.4 & 3.5 & 2.9 \\
$M_D$ for $(P_{e^-},P_{e^+})=(80\,\%,60\,\%)$   
                                           & 10.4 & 6.9 & 5.1 & 4.0 & 3.3 \\ 
\end{tabular}
\end{table}

\section*{Acknowledgements}

A.B.\ is grateful to the organisers of WHEPP-8 for their invitation
and hospitality and for creating an inspiring atmosphere.
This work was supported by the `Fonds zur F\"orderung der
wissenschaftlichen For\-schung' of Austria, FWF Project No.~P16592-N02
and by the European Community's Human Potential Programme
under contract HPRN-CT-2000-00149.

\end{document}